\documentclass[aps,prd,twocolumn,noshowpacs,nofootinbib,superscriptaddress,groupedaddress, floatfix,superscriptaddress, preprintnumbers]{revtex4-2}

\usepackage[utf8]{inputenc}
\usepackage{dcolumn}
\usepackage{amsmath}
\usepackage{amsfonts}
\usepackage{graphicx}
\usepackage{amssymb}%
\usepackage{array,multirow}
\usepackage[breaklinks,colorlinks=true]{hyperref}
\begin{document}

\begin{flushright}
\end{flushright}


\title{Invisible Charm Exotica}
\author{Micha\l~Prasza\l owicz}
\email{michal.praszalowicz@uj.edu.pl}
\affiliation{Institute of Theoretical Physics, Jagiellonian University, S. \L ojasiewicza 11, 30-348 Krak\'ow, Poland}

\author{Maciej Kucab}
\email{maciej.kucab@student.uj.edu.pl}
\affiliation{Institute of Theoretical Physics, Jagiellonian University, S. \L ojasiewicza 11, 30-348 Krak\'ow, Poland}

\date{\today}

\begin{abstract}
  One possible interpretation of two narrow  $\Omega^0_c$ states reported by the LHCb Collaboration at CERN 
  in 2017 is that they are pentaquarks belonging to a $\overline{\boldsymbol{15}}$ exotic SU(3) representation,
  as predicted by the Chiral Quark--Soliton Model. If so, there must exist a number of other exotic states since
  the model predicts three different $\overline{\boldsymbol{15}}$ multiplets of heavy baryons. We show
  that, depending on the soliton spin $J$, these states
  are either very narrow or very broad. This explains why they might have escaped experimental observation.
  Furthermore, we show that the lightest members of these multiplets are stable against two body strong decays.  
  \end{abstract}

\maketitle

\section{Introduction}

Heavy baryon spectroscopy has recently attracted attention triggered by the discoveries of new states including
hidden charm pentaquarks and tetraquarks. Present situation in the charm sector
has been recently reviewed in Ref.~\cite{Cheng:2021qpd}.
Here, in this paper, we will concentrate on heavy baryons with one charm quark. These states can be conveniently
classified according to the SU(3) structure of the light quarks, which can form an antitriplet of spin 0 or a sextet of spin 1.
Adding a charm quark results in an antitriplet of spin 1/2 and two hyperfine split sextets of spin 1/2 and 3/2. This structure
is fully confirmed by experiment \cite{PDG}.

It was pointed out in Refs.~\cite{Yang:2016qdz,Kim:2017jpx,Kim:2017khv,Polyakov:2022eub} that exactly the same SU(3)
structure follows from the Chiral Quark--Soliton Model ($\chi$QSM)
as a result of the quantization of the soliton rotations. At the same time, higher rotational excitations
have been shown to correspond to the exotic baryons -- pentaquarks \cite{Praszalowicz:2003ik,Diakonov:1997mm}.
In the present context, the lowest lying exotic SU(3) representation is $\overline{\boldsymbol{15}}$ \cite{Kim:2017jpx}.

In the quark model, one of the  possible excitations consists in adding  angular momentum, 
which in the heavy quark rest frame  may be interpreted as the angular momentum of
the light quarks. Such a configuration has negative parity. An immediate consequence  
of this picture  is the emergence of two hyperfine
split antitriplets of spin $1/2^-$  and $3/2^-$ that are indeed observed experimentally both in charm
and (partially) bottom sectors. In the sextet case, the total angular momentum of the light subsystem can be 0, 1, or 2.
Therefore, one predicts five excited sextets of negative parity: two with total spin $1/2$, two with total
spin 3/2, and one with total spin 5/2. Again the same structure is predicted by the $\chi$QSM \cite{Polyakov:2022eub}.

 In 2017, the LHCb Collaboration
announced five new excited $\Omega_{c}^{0}$ states, two of them of a very
small width~\cite{LHCb:2017uwr}, which were confirmed by the Belle
Collaboration~\cite{Belle:2017ext} in 2018. Further analysis of the decay
modes and possible spin assignment of these states has been published 
recently in Ref.~\cite{LHCb:2021ptx}.

The  LHCb resonances could be the first experimentally observed particles from the
negative parity excited sextet. Such an assignment has been advocated in 
Refs.~\cite{Karliner:2017kfm,Wang:2017vnc,Chen:2017gnu,Santopinto:2018ljf,Jia:2020vek}
in different versions of the quark model, within the QCD sum rules \cite{Wang:2017zjw}
and lattice QCD \cite{Padmanath:2017lng}.

Unfortunately, when it comes to a more detailed  analysis of the LHCb
data, basically all the models have problems to accommodate all five LHCb resonances within the above scenario
with acceptable accuracy. Therefore alternative assignments of some of the LHCb resonances have been proposed.
The comprehensive summary of different assignments can be found in Sec.~II.3 of the recent review by Cheng~\cite{Cheng:2021qpd}.

In Ref.~\cite{Kim:2017jpx}  two
narrow LHCb $\Omega^0_c$ states, namely $\Omega_{c}^{0}(3050)$ and $\Omega_{c}^{0}(3119)$,
have been interpreted 
as the hyperfine split members of the exotic $\overline{\boldsymbol{15}}$. This
assignment has been motivated by the fact\footnote{Note that the ground state
sextet and exotic $\overline{\boldsymbol{15}}$ belong to the same rotational band, and
therefore should have approximately the same value of the hyperfine splitting.} that their
hyperfine splitting is equal to the one of the ground state sextet and has been further reinforced by
the calculation of their widths~\cite{Kim:2017khv}. Alternative pentaquark interpretations
can be found in Refs.~\cite{An:2017lwg,Yang:2017rpg,Wang:2018alb,Wang:2017smo}.

Introducing new exotic multiplets, in itself very attractive, is nevertheless a phenomenological challenge. 
In fact we have two exotic $\overline{\boldsymbol{15}}$ light SU(3) multiplets. One, carrying angular momentum $J=1$, leads
to two hyperfine split heavy baryon multiplets, and the second one with $J=0$
corresponds to yet another heavier multiplet, whose properties 
so far have not been  discussed in detail~\cite {Kim:2017jpx}; see however model calculations
of Ref.~\cite{Kim:2019rcx}. 
So we have introduced 45 new particles
(or perhaps it is better to say: 18 isospin submultiplets), out of which only two $\Omega^0_c$ states
(members of two isospin multiplets) have been  used in phenomenology. Where are the remaining
states?

To answer this question we compute in the framework of the $\chi$QSM masses and strong  decay widths 
of all these supernumerary states. We find that the members of the multiplets based on ($\overline{\boldsymbol{15}}$,
$J=1$) soliton are very narrow (some hint of this behavior has been already discussed in Ref.~\cite{Kim:2017khv}),
and -- on the contrary -- states associated with ($\overline{\boldsymbol{15}}, J=0)$ multiplet are  wide.
Both extremes explain why these states have not been seen experimentally: it is easy to overlook a very narrow or
very broad resonant signal.
We also find that the nucleonlike isospin dublet of $J=1$ and also $J=0$ soliton seems to be stable against 
two body strong decays.

In order to compute masses and decay widths one has to fix model parameters. In 
Refs.~\cite{Yang:2016qdz,Kim:2017jpx,Kim:2017khv}, these parameters have been fixed
from the phenomenology of the light baryons, with a modification  based on the $N_c$ counting.
In Ref.~\cite{Kim:2019rcx}, they have been computed in a specific model.
Here, we follow a different strategy, namely we fix mass parameters from the heavy baryon sector alone.
Predicted masses are in agreement with Ref.~\cite{Kim:2017jpx}. For decays, we use parameter
values from Ref.~\cite{Kim:2017khv}.

The paper is organized as follows. In Sec.~\ref{sec:chiQSM} we briefly review the main features of the $\chi$QSM.
In Sec.~\ref{sec:HBmasses} we first derive analytical formulae for the baryon masses and then fix splitting
 parameters as functions of the strange moment of inertia $1/I_2$. After constraining $1/I_2$ we compute all pentaquark 
 masses. Next, in Sec.~\ref{sec:decays}, we discuss and compute decays widths, and finally we conclude in Sec.~\ref{sec:conclusions}.

\section{Chiral Quark--Soliton Model}

\label{sec:chiQSM}

In this section, we briefly recapitulate the main features of the $\chi$QSM that can be found
in the original paper by Diakonov, Petrov and Pobylitsa
\cite{Diakonov:1987ty} and in the reviews of
Refs.~\cite{Christov:1995vm,Alkofer:1994ph,Petrov:2016vvl} (and references
therein). The $\chi$QSM is based  on the large $N_{c}$  argument by
Witten~\cite{Witten:1979kh,WittenCA}, which says that for
$N_{c} \rightarrow\infty$,  $N_{c}$ relativistic valence quarks generate
chiral mean fields represented by a distortion of  the Dirac sea. This
distortion in turn interacts with the valence quarks,
which in turn modify the sea until a stable configuration is reached. 
Such a
configuration is called   \emph{chiral soliton}. It is a solution of the
Dirac equation for the constituent quarks (with gluons integrated out) in the
mean-field approximation. 

The soliton does not carry
any quantum numbers except for the baryon number resulting from the
valence quarks. Spin and isospin appear when the soliton rotations in space and flavor are quantized~\cite{Adkins:1983ya}.
This procedure results in a {\em collective} Hamiltonian analogous to the one
of a quantum mechanical symmetric top, however, due to the Wess-Zumino-Witten
term \cite{WittenCA,Wess:1971yu}
the allowed Hilbert space is truncated to the representations that contain states of
hypercharge $Y'=N_{\rm val}/3$. For $N_{\rm val}=3$, these are an octet and decuplet of ground state
baryons~\cite{Guadagnini:1983uv,Mazur:1984yf,Jain:1984gp} .

In order describe heavy baryons we have to remove one quark from the valence level and replace it by
a heavy quark $Q$. Formally, this corresponds to a replacement of $N_c$  light  valence quarks by $N_c-1$ quarks.
In the limit $N_c \rightarrow \infty$ such a replacement does not parametrically change the mean fields; however, for $N_c=3$, we should
expect that the numerics of the model will be modified. Moreover, the isospin $T'$ of the states
with a hypercharge equal to $Y'$ is equal to the soliton angular momentum~\cite{Guadagnini:1983uv,Mazur:1984yf,Jain:1984gp}, 
which in the following
will be denoted by $J$.

\begin{figure}[h]
\centering
\includegraphics[width=9cm]{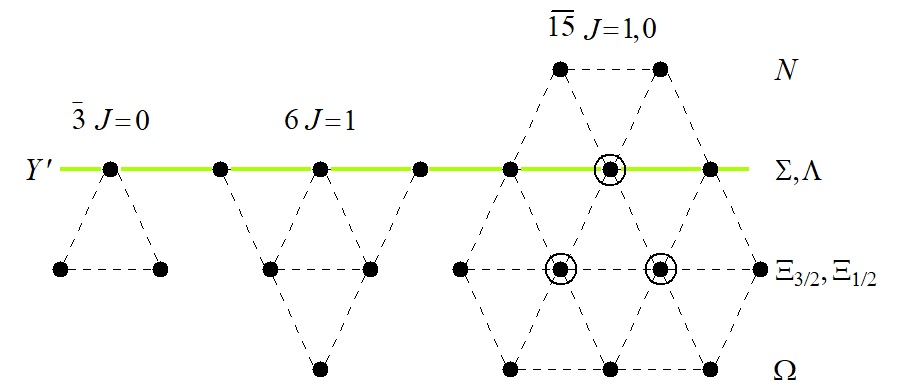} \vspace{-0.2cm}%
\caption{Rotational band of a soliton with one valence quark stripped off.
Soliton spin corresponds to the isospin $T^{\prime}$ of states on the
quantization line $Y^{\prime}=2/3$ (green thick line). We show three lowest
allowed representations: antitriplet of spin 0, sextet of spin 1, and the
lowest exotic representation $\overline{\mathbf{{15}}}$ of spin 1 or 0. 
On the right-hand side we display particle names used
in the present paper.}%
\label{fig:reps}%
\end{figure}

In this picture, the allowed SU(3) representations  have to contain states of $Y'=2/3$, and 
these are $\overline{\boldsymbol{3}}$, $\boldsymbol{6}$, and exotic
$\overline{\boldsymbol{15}}$ shown in Fig.~\ref{fig:reps}. They correspond to the
rotational excitations of the meson mean field, which is essentially the same
as for light baryons. The corresponding wave function of the light sector is
given in terms of the Wigner rotational $D(A)$ matrices~\cite{Diakonov:1997mm}
\begin{align}
\psi_{(\overline{\mathcal{R}}\,;\,-Y^{\prime}\,J\,J_{3})}^{(\mathcal{R\,};\,B)}(A)  
 = & \sqrt{\text{dim}(\mathcal{R})}\,(-)^{J_{3}-Y^{\prime}/2} \notag \\
 \times&D_{(Y,\,T,\,T_{3}%
)(Y^{\prime},\,J,\,-J_{3})}^{(\mathcal{R})\ast}(A) 
\label{eq:rotwf}%
\end{align}
where $\mathcal{R}$ denotes the SU(3) representation of the light sector,
$B=(Y,T,T_{3})$ stands for the SU(3) quantum numbers of a baryon in question,
and the second index of the $D$ function, $(Y^{\prime},J,-J_{3})$, corresponds
to the soliton spin.  $A(t)$
denotes relative \emph{configuration space} -- SU(3) \emph{group space}
rotation matrix.

The total wave function of a heavy baryon of spin $s$ is constructed by
coupling (\ref{eq:rotwf}) to a heavy quark spinor $\chi_{M}^{1/2}$
with a pertinent SU(2) Clebsch-Gordan coefficient,
\begin{equation}
\Psi_{B\, J\,s\,m}^{(\mathcal{R})}=\sum_{M,M_{J}}\chi_{M}^{1/2}\otimes \,
\psi_{(\mathcal{\bar{R}};\,-Y^{\prime}\,J\,M_{J})}^{(\mathcal{R\,}%
;\,B)} \; \left(
\begin{array}
[c]{cc}%
J & 1/2\\
M_{J} & M
\end{array}
\right\vert \left.
\begin{array}
[c]{c}%
s\\
m
\end{array}
\right)  .
\label{eq:fullwf}
\end{equation}

The rotational Hamiltonian takes the following form~\cite{Diakonov:2013qta} 
in the chiral limit:
\begin{align}
\mathcal{M}_{\mathcal{R}}=& M_{\text{sol}}+\;\frac{1}{2I_{2}}\left[
C_{2}(\mathcal{R})-T^{\prime}(T^{\prime}+1)-\frac{3}{4}Y^{\prime2}\right] \notag \\
+& \frac{1}{2I_{1}}T^{\prime}(T^{\prime}+1) \label{rotmass}%
\end{align}
where $C_{2}(\mathcal{R})$ stands for the SU(3) Casimir operator.
$M_{\text{sol}}\sim N_{c}$ denotes the classical soliton mass; $I_{1,2}\sim N_{c}$
are moments of inertia. All these parameters can, in
principle, be calculated in some specific model. Here, we shall follow a so-called
{model-independent} approach introduced in the context of the Skyrme
model in Ref.~\cite{Adkins:1984cf}, where all  parameters are extracted from
the experimental data~\cite{Kim:2017jpx}.

The symmetry
breaking Hamiltonian takes the following form~\cite{Blotz:1992pw}:
\begin{equation}
H_{\mathrm{{sb}}}=\alpha\,D_{88}^{(8)}+\beta\,\hat{Y}+\frac{\gamma}{\sqrt{3}%
}\sum_{i=1}^{3}D_{8i}^{(8)}\,\hat{J}_{i}, \label{eq:Hsb}%
\end{equation}
where $\alpha$, $\beta$, and $\gamma$ are proportional to the strange
quark mass and are
given in terms of the moments of
inertia and the pion-nucleon sigma term. Their explicit form is not of
interest to us, as we shall treat them as free parameters. It is however worth
mentioning that $\alpha$ and $\beta$ are negative by construction, while
$\gamma$ being phenomenologically negative is in fact given as a difference of
two terms of the same order -- see Eq.~(4) in Ref~\cite{Yang:2016qdz}. Furthermore,
$\alpha$  scales as $N_c$, and $\beta$ and $\gamma$
scale as $N_c^0$.

The soliton of $J=1$ can couple with the heavy quark to
baryon spin  $s=1/2$ or $s=3/2$. 
These states will be hyperfine
split, and in order to take this into account, following \cite{Yang:2016qdz}, 
we supplement Hamiltonians (\ref{rotmass}) and (\ref{eq:Hsb})
with
 the chromomagnetic interaction~\cite{Yang:2016qdz}
expressed as:
\begin{align}
H_{SQ} = \frac{2}{3}\frac{\varkappa}{m_{Q}} \hat{J} \cdot
\hat{S}_{Q} \label{eq:ssinter}%
\end{align}
where $\varkappa$ denotes the anomalous chromomagnetic moment that is
flavor independent. The operators ${\hat{J}}$ and ${\hat{S}}_{Q}$
represent the spin operators for the soliton and the heavy quark,
respectively.

\section{Masses of heavy baryons}
\label{sec:HBmasses}

\subsection{General formulas}

As we can see from Fig.~\ref{fig:reps}, the soliton in $\overline
{\mathbf{{15}}}=(p=1,q=2)$ can be quantized both as spin $J=0$ and $1$ (remember that
the isospin of the states on $Y^{\prime}=2/3$ line corresponds to
spin\footnote{From now on we use numerical values of the quantum numbers
corresponding to $N_{c}=3$, which does not allow for proper $N_{c}$ counting.}).
 Next
possible exotic representation is $\overline{\mathbf{15}}^{\prime}=(p=0,q=4)$
with spin $J=1$, which however, is heavier than $\overline{\mathbf{{15}}}$.

In order to estimate the masses of states in $\overline{\mathbf{{15}}}$ we
shall use the general formula (\ref{rotmass}) for the rotational energy of the
soliton.
For the exotic representations in question we have%
\begin{eqnarray}
\mathcal{M}_{\overline{\mathbf{{15}}},J=0}&=&M_{\mathrm{{{sol}}}}+\frac{5}%
{2}\frac{1}{I_{2}}\, , \nonumber \\
\mathcal{M}_{\overline{\mathbf{{15}}},J=1}%
&=&M_{\mathrm{{{sol}}}}+\frac{3}{2}\frac{1}{I_{2}}+\frac{1}{I_{1}}.
\label{eq:M15bar}%
\end{eqnarray}
Interestingly, the mass difference
\begin{eqnarray}
\Delta_{\overline{\mathbf{{15}}}}&=&\mathcal{M}_{\overline{\mathbf{{15}}}%
,J=0}-\mathcal{M}_{\overline{\mathbf{{15}}},J=1}=\frac{1}{I_{2}}-\frac
{1}{I_{1}}%
\label{eq:hfsplit}
\end{eqnarray}
is expected to be positive, since -- from the estimates of the light sector
\cite{Kim:2017jpx,Diakonov:1997mm} -- $I_{1}\sim(2.5\div3)\times I_{2}$, 
which means that spin 1 soliton is
lighter than the one of spin 0. One of the goals of the present analysis is to
constrain these two parameters from the heavy sector alone. Indeed, solitons
considered here are constructed from $N_{c}-1$ valence quarks, what may
finally result in a change of the numerical values of $I_{1,2}$ as compared to
the values extracted from the light sector~\cite{Kim:2017jpx}.

The average multiplet masses take the following form:%
\begin{align}
M_{\overline{\mathbf{{3}}},J=0}^{Q}  &  =m_{Q}+M_{\mathrm{{sol}}}+\frac{1}{2I_{2}%
},\nonumber\\
M_{\mathbf{{6}},J=1}^{Q}  &  =M_{\overline{\mathbf{{3}}}}^{Q}+\frac{1}{I_{1}%
},\nonumber\\
M_{\overline{\boldsymbol{15}},J=1}^{Q}  &  =M_{\mathbf{{6}}}^{Q}+\frac
{1}{I_{2}},\nonumber\\
M_{\overline{\boldsymbol{15}},J=0}^{Q}  &  =M_{\mathbf{{6}}}^{Q}+\frac
{2}{I_{2}}-\frac{1}{I_{1}}=M_{\overline{\boldsymbol{15}},J=1}^{Q}%
+\Delta_{\overline{\mathbf{{15}}}}. \label{eq:15barav}%
\end{align}
Parameters $M_{\overline{\mathbf{{3}}}}^{Q}$ and $I_{1}$ can be extracted from
the ground state nonexotic baryons~\cite{Yang:2016qdz}. In order to have some handle on $I_{2}$,
and therefore on $\Delta_{\overline{\mathbf{{15}}}}$,
we shall include now flavor symmetry breaking due to the mass difference
between strange and non-strange quarks (\ref{eq:Hsb}). 

Calculating matrix elements of the symmetry breaking operator  (\ref{eq:Hsb}) between the
collective wave functions (\ref{eq:rotwf}) we obtain the following mass splittings for the ground
state and excited baryons:
\begin{align}
\left\langle H_{\text{sb}}\right\rangle _{\overline{\boldsymbol{3}},J=0}  &
=\left(\frac{3}{8}{\alpha}+\beta \right) Y_B = \delta_{\overline{\mathbf{{3}}}} Y_B\, , \notag \\
\left\langle H_{\text{sb}}\right\rangle _{{\boldsymbol{6}},J=1}  &
=\left( \frac{3}{20}{\alpha}+\beta-\frac{3}{10}\gamma\right) Y_B= \delta_{\mathbf{{6}}} Y_B\, , \notag \\
\left\langle H_{\text{sb}}\right\rangle _{\overline{\boldsymbol{15}},J=1}  &
=\left(  \beta+\frac{17}{144}(\alpha-2\gamma)\right)  Y_{B} - (\alpha-2\gamma) \notag \\
&\times \left(  \frac{2}{27}-\frac{1}{24}\left(  T_{B}(T_{B}+1)-\frac{1}{4}Y_{B}^{2}\right)
\right)  \,,\nonumber\\
\left\langle H_{\text{sb}}\right\rangle _{\overline{\boldsymbol{15}},J=0}  &
=\left(  \beta+\frac{1}{48}\alpha\right)  Y_{B} \notag \\
&+\alpha \left(  \frac{2}{9}-\frac
{1}{8}\left(  T_{B}(T_{B}+1)-\frac{1}{4}Y_{B}^{2}\right)  \right)  \,
\label{eq:Delta15bar}%
\end{align}
where $Y_{B}$ and $T_B$  denote the hypercharge and the isospin of a given baryon,
respectively. In the case of
sextet and  (${\overline{\boldsymbol{15}},J=1}$), the mass formula  must be supplemented by the spin
splitting Hamiltonian (\ref{eq:ssinter}), leading to the following equations for baryon masses
\begin{align}
M_{\mathcal{R}_J,B,s}^{Q}&=M_{\mathcal{R},J}^{Q}+\left\langle H_{\text{sb}}\right\rangle _{\mathcal{R}, J} \notag \\
&+\delta_{J,1} 
\frac{\varkappa}{m_{Q}}\left\{
\begin{array}
[c]{ccc}%
-2/3 & \text{for} & s=1/2\\
\, & \, & \,\\
+1/3 & \text{for} & s=3/2
\end{array}
\right.
\label{eq:M3barM6mass}%
\end{align}
where $s$ denotes the spin of a given baryon, and $J$ is the soliton spin.
It is worth to observe that the hyperfine splitting parameter $\varkappa/m_Q$
can estimated from the following mass differences:
\begin{align}
M_{\mathcal{R}_1,B,3/2}^{Q}\;-\; M_{\mathcal{R}_1,B,1/2}^{Q} = \frac
{\varkappa}{m_{Q}} . \label{eq:DCsextet}%
\end{align}

It turns out that the mass formulas (\ref{eq:Delta15bar}) for $\overline
{\boldsymbol{15}}$ admit three Gell-Mann--Okubo (GMO)~\cite{Gell-Mann:1962yej,Okubo:1961jc}
 mass
relations,\footnote{Whenever this does not cause confusion, we use particle
symbols to denote their masses.}%
\begin{align}
2N^{Q}+\Omega^{Q}  &  =2\Lambda^{Q}+\Sigma^{Q},\nonumber\\
N^{Q}+\Xi^{Q}_{3/2}  &  =2\Sigma^{Q},\nonumber\\
2N^{Q}+2\Xi^{Q}_{1/2}  &  =3\Lambda^{Q}+\Sigma^{Q}, \label{eq:GMO}%
\end{align}
both in $J=1$ and $J=0$ multiplets. Although the mass formulas for both
multiplets differ, the GMO mass relations are identical. It might be at the first
sight surprising that for six isospin multiplets whose masses in the case of
$\overline{\boldsymbol{15}}_{J=1}$ are parametrized by four parameters:
$M^{Q}_{\overline{\boldsymbol{15}},J=1},\alpha,\beta$ and $\gamma$ we have
three sum rules rather than two. The reason is that the splittings depend only
on the combination $\alpha-2\gamma$. Relations (\ref{eq:GMO}) are linearly
independent but not orthogonal. Furthermore, the following Guadagnini-type
relation~\cite{Guadagnini:1983uv} is fulfilled:%
\begin{align}
& \left[  21N^{Q}-2\Lambda^{Q}+16\Sigma^{Q}-16\Xi^{Q}_{1/2}+11\Xi^{Q}%
_{3/2}-30\Omega^{Q}\right]  _{\overline{\boldsymbol{15}}_0}  
\notag \\
=&\left[  30N^{Q}+34\Lambda^{Q}-14\Sigma^{Q}+14\Xi^{Q}_{1/2}-58\Xi^{Q}%
_{3/2}-6\Omega^{Q}\right]  _{\overline{\boldsymbol{15}}_1}. \notag \\
&~~
 \label{eq:Gaud}%
\end{align}
Relation (\ref{eq:Gaud}) has been constructed by demanding orthogonality to
relations (\ref{eq:GMO}). It connects masses of different multiplets and therefore
goes beyond the SU(3) symmetry.

\subsection{Numerical estimates}

Let us first consider masses of the nonexotic heavy baryons belonging to
$\boldsymbol{\bar{3}}$ and $\boldsymbol{6}$ of SU(3). The average masses of
these multiplets are given by Eqs.~(\ref{eq:15barav}), in fact both for $Q=c$ and $b$,
\begin{align}
M_{\overline{\mathbf{{3}}}}^{Q}=m_{Q}+M_{\mathrm{{sol}}}+\frac{1}{2I_{2}}  &
=\left.  2408.2\right\vert _{c}~,~\left.  5736.2\right\vert _{b}\,,\nonumber\\
M_{\mathbf{{6}}}^{Q}=M_{\overline{\mathbf{{3}}}}^{Q}+\frac{1}{I_{1}}  &
=\left.  2579.7\right\vert _{c},~~\left.  5906.5\right\vert _{b} \label{eq:mqrep}%
\end{align}
where the experimental values in MeV from Ref.~\cite{Yang:2016qdz} have been updated~\cite{Praszalowicz:2022sqx}. We can
compute $I_{1}$ from the mass difference of these two multiplets (in MeV):

\begin{equation}
\frac{1}{I_{1}}=M_{\mathbf{6}}^{Q}-M_{\mathbf{\overline{3}}}^{Q}=\left.
171.5\right\vert _{c}=\left.  170.4\right\vert _{b} 
\label{eq:I1Q}%
\end{equation}

Similarly we can compute heavy quark mass difference either from the mass
difference of the bottom or charm antitriplets or sextets,
\begin{align}
m_{b}-m_{c}=  &  M_{\mathbf{6}}^{b}-M_{\mathbf{{6}}}^{c}=3327~\mathrm{{MeV}%
,}\nonumber\\
=  &  M_{\mathbf{\overline{3}}}^{b}-M_{\mathbf{\overline{3}}}^{c}%
=3328~\mathrm{{MeV}. } \label{eq:mbmmc}%
\end{align}

We consider perfect equality of splittings (\ref{eq:I1Q}) regardless of $Q$ 
and the mass difference (\ref{eq:mbmmc})
regardless of the SU(3) representation, as a test of our model assumptions.
Equalities (\ref{eq:I1Q}) and (\ref{eq:mbmmc}) can be traced back to the fact that in the present model heavy baryon mass is
simply a sum of a heavy quark mass and the rotational excitations of the
soliton, see  Eqs.~(\ref{eq:15barav}), which are flavor-blind in the present approach. Moreover, the effects
of  SU(3) symmetry breaking are simply the same both for
charm and bottom baryons, since they are solely due to the 
light quarks within the soliton.

Numerical value of $1/I_1$ from Eq.~(\ref{eq:I1Q}) 
should be compared with $1/I_{1}$ extracted from the light
sector, which is equal to $\sim155~{\rm MeV}$ \cite{Ellis:2004uz}. This is consistent with the
expectation that moments of inertia should be smaller in the case of heavy
baryons, since the valence quark contributions to $I_{1,2}$ scales like
$N_{\text{val}}$. In what follows, we shall assume $1/I_{1}=171$~MeV.
Unfortunately, we cannot extract $I_{2}$ in a model independent way from the
masses of the ground state multiplets. To this end, we have to use information
from the mass splittings within different multiplets, including
exotica.

In Ref.~\cite{Yang:2016qdz} the splitting parameters for $\overline{\boldsymbol{3}}$ 
and $\boldsymbol{6}$ have been extracted from experiment and read
\begin{align}
\delta_{\overline{\mathbf{{3}}}}=\frac{3}{8}{\alpha}+\beta &
=-180~\mathrm{{MeV}\,,}\nonumber\\
\delta_{\mathbf{{6}}}=\frac{3}{20}{\alpha}+\beta-\frac{3}{10}\gamma &
=-121~\mathrm{{MeV}\,. } \label{eq:d3bard6}%
\end{align}
Numerical entries are taken as the average values from Eqs.~(13) and (14) in Ref.~\cite{Yang:2016qdz}.

\begin{figure}[h]
\centering
\includegraphics[width=7cm]{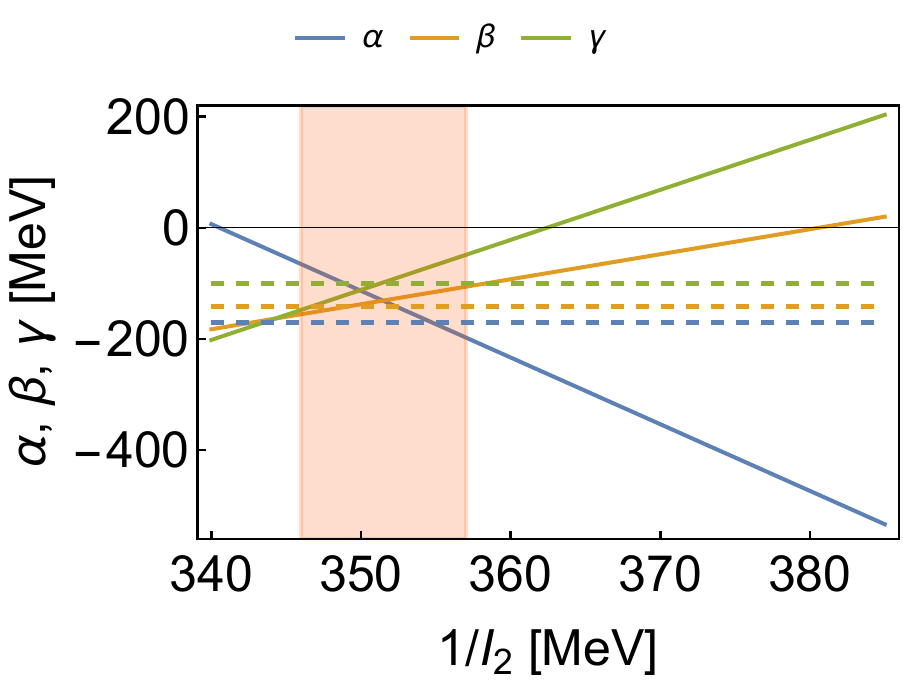} \vspace{-0.2cm} \caption{Parameters
$\alpha$, $\beta$, and $\gamma$ plotted in terms of the inverse moment of
inertia $1/I_{2}$. Expectations from the light sector are shown as dashed
lines.}%
\label{fig:albegam2}%
\end{figure}

In Ref.~\cite{Kim:2017jpx}, two out of five excited $\Omega_{c}$ hyperons reported by the
LHCb Collaboration in 2017 \cite{LHCb:2017uwr} have been interpreted as exotic states
belonging to $(\overline{\mathbf{{15}}},J=1)$. Adding a heavy quark to the
$J=1$ soliton results in two hyperfine split states (\ref{eq:M3barM6mass}) of spin
$1/2$ and $3/2$ , namely $\Omega_{c}(3050)$ and $\Omega_{c}(3119)$,
respectively. This splitting (\ref{eq:DCsextet}) is equal to $\varkappa
/m_{c}=69$~MeV \cite{Yang:2016qdz,Kim:2017jpx}. $\Omega_{c}$ average mass before the
spin splitting is%
\begin{equation}
\overline{M}_{\Omega,(\overline{\mathbf{{15}}},J=1)}^{c}=3096\;\text{MeV.}
\label{eq:MavOmega}%
\end{equation}
From Eqs.\thinspace(\ref{eq:M15bar}), (\ref{eq:15barav}) and (\ref{eq:Delta15bar}) 
we obtain that%
\begin{align}
\overline{M}_{\Omega,(\overline{\mathbf{{15}}},J=1)}^{c}  &  
=M_{\mathbf{{6}}}^{c}+\frac{1}{I_{2}}-\frac{1}{6}\left(  \alpha
+8\beta-2\gamma\right)  \label{eq:MavOmegath}%
\end{align}
Equating
(\ref{eq:MavOmega}) with (\ref{eq:MavOmegath}) together with
Eqs.~(\ref{eq:d3bard6}) gives three independent equations for four parameters
$\alpha$, $\beta$, $\gamma$ and $1/I_{2}$. We solve them in function of
$1/I_{2}$ and constrain parameter $1/I_{2}$ to the region where both $\alpha$
and $\beta$ are negative. The result is plotted in Fig.~\ref{fig:albegam2}.

We see from Fig.~\ref{fig:albegam2} that the allowed range (\emph{i.e.}, the
range where $\alpha,\beta<0$) for the second moment of inertia is
$342~\mathrm{{MeV}<1/I_{2}<380}$~MeV. However, all model calculations and fits
to the light sector suggest that also parameter $\gamma$ should be negative (see, e.g., Ref.~\cite{Kim:2019rcx});
then the allowed range for $1/I_{2}$ is further constrained to $1/I_{2}%
<363$~MeV. The most probable value of $1/I_{2}$ is around 351~MeV, where all
splitting parameters are negative and of the same order. Indeed, from fits to
the light sector one obtains \cite{Yang:2016qdz} $\alpha= -255~\mathrm{{MeV},
\;\;\; \beta= -140~{MeV}, \;\;\; \gamma= -101~{MeV}}$. However, as explained
in Ref.~\cite{Yang:2016qdz}, the parameter $\alpha$ scales linearly with the number of valence
quarks, $N_{\mathrm{{val}}}$, whereas parameters $\beta$ and $\gamma$ are in
the first approximation independent of $N_{\mathrm{{val}}}$, because they are
equal to the ratios of quantities that scale like $N_{\mathrm{{val}}}$. This
means that in the heavy baryon sector, we expect $\alpha\rightarrow
2/3\times(-255)= -170$~MeV. From our fits for $1/I_{2}\simeq351$~MeV, we obtain
$\alpha\simeq-110$~MeV, $\beta\simeq-139$~MeV and $\gamma\simeq-114$~MeV. Here
only $\alpha$ is substantially different from the light sector estimate. This
is shown in more detail in Fig.~\ref{fig:albegam2} where model expectations
from the light sector are shown as dashed lines. In what follows, we shall
discuss the sensitivity of heavy pentaquark masses to the variation of
$1/I_{2}$ within the limits $\pm5$~MeV around 351~MeV. This is shown as a
light-orange band in Fig.~\ref{fig:albegam2}.

Finally, let us observe that assuming $1/I_{2}=351$~MeV and taking $1/I_{1}$
from Eq.~(\ref{eq:I1Q}), we obtain that $(\overline{\mathbf{{15}}},J=0)$
multiplet is heavier from $(\overline{\mathbf{{15}}},J=1)$ multiplet on
average by approximately $180$~MeV. We have, therefore,%
\begin{align}
M_{\overline{\boldsymbol{15}},J=1}^{c}  &  \simeq2931\;\text{MeV},\nonumber\\
M_{\overline{\boldsymbol{15}},J=0}^{c}  &  \simeq3111\;\text{MeV.}%
\end{align}
At this point, we can estimate the average mass of the next exotic
representation $(\overline{\boldsymbol{15}}^{\prime}=(p=0,q=4),J=1)$ to be
approximately 3633 MeV, which is indeed substantially heavier than
$\overline{\boldsymbol{15}}$.

\renewcommand{\arraystretch}{1.7}
\begin{table}[h!]
\centering
\begin{tabular}{|c|c|c|c|}
\hline
\multirow{2}{*}{$\overline{\boldsymbol{15}}$}& \multicolumn{2}{c|}{$J=1$} &$ J=0$\\
\cline{2-4}
&$ s=1/2$ &$ s=3/2$ & $s=1/2$\\
\hline
$N^{c} $& 2644--2692 & 2713--2761 & 2819--2884\\
$\Lambda^{c}$ & 2772--2812 & 2841--2881 & 2981--3001\\
$\Sigma^{c}$ & 2795--2810 & 2864--2879 &2993--3043 \\
$\Xi_{1/2}^{c} $& 2911--2931 & 2980--3000 & 3148--3138\\
$\Xi_{3/2}^{c}$ & 2945--2927 & 3014--2996 & 3167--3202\\
$\Omega^{c}$ & 3050 & 3119 & 3316--3276 \\
\hline
\end{tabular}
\caption{Mass predictions in MeV for exotic $\overline{\boldsymbol{15}}$.
Two $\Omega^{c}$ states are taken as input.
\label{tab:mpred}}
\end{table}
\renewcommand{\arraystretch}{1}

\begin{figure*}[t]
\centering
\includegraphics[width=14cm]{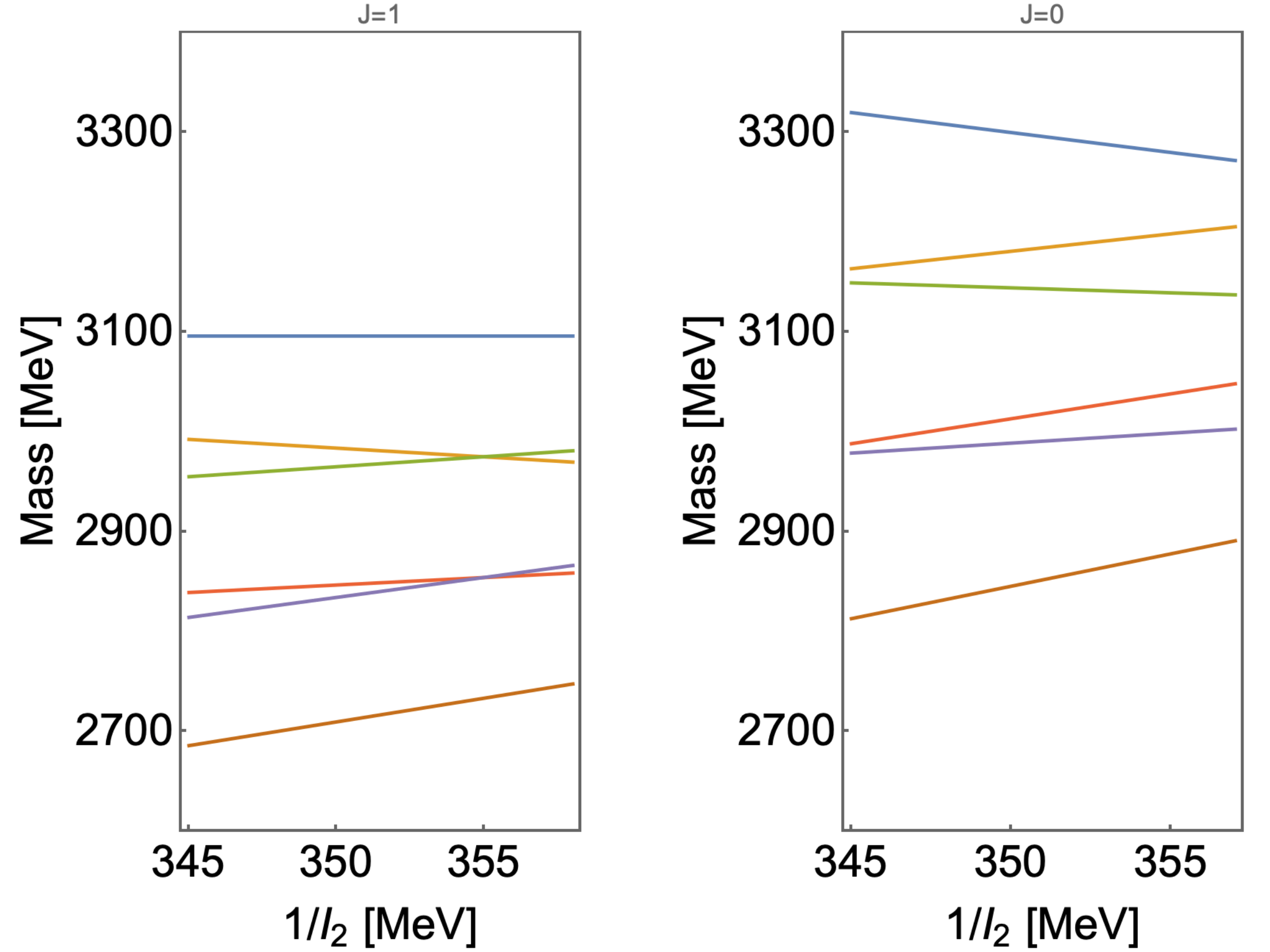} 
\caption{Spectra of exotic charm multiplets
${\overline{\boldsymbol{15}},J=1}$ (left) 
and $J=0$
(right) 
in terms of the inverse moment of inertia
$1/I_{2}$.}%
\label{fig:spectra}%
\end{figure*}

In Fig.~\ref{fig:spectra} and in Tab.~\ref{tab:mpred} we show the results
for $\overline{\boldsymbol{15}}$ masses both for $J=$ and $J=0$.
Our predictions for  $J=1$ multiplets are in agreement with Ref.~\cite{Kim:2017jpx},
where parameters $\alpha$, $\beta$, and $\gamma$ have been estimated from
the light sector alone.

It is interesting to compare our phenomenological results with model calculations of Ref.~\cite{Kim:2019rcx}.
Using modified chiral fields, they obtain $1/I_2 \simeq 380$~MeV, i.e. above the upper edge of our allowed window
shown in Fig~\ref{fig:albegam2}.
This is the reason why their mass predictions are higher than in the present work. Their parameters
$\alpha$ and $\beta$ are similar to ours, although $\alpha$ is a bit smaller and $\beta$ a bit larger.
On the other hand, $\gamma$ is much larger than in our case, but still negative.
The latter explains why their value of $\delta_{\boldsymbol{6}}$ undershoots experiment (\ref{eq:d3bard6})
by $\sim 15\%$.
The values of their parameters are, however, consistent
with the dependence on $1/I_2$ displayed in Fig~\ref{fig:albegam2}. It should be stressed
that the calculations in Ref.~\cite{Kim:2019rcx} have been done for one particular choice of chiral fields,
namely for the pseudoscalars only.
Nevertheless,  their results support our initial conjecture that for $N_c=3$, we expect that numerical values
of model parameters differ depending of the number of  quarks in the soliton valence level.

\section{Decays}
\label{sec:decays}
\subsection{General formulas}

To calculate the decays of  heavy baryons, one has to sandwich the
corresponding decay operator between the wave functions (\ref{eq:fullwf}).
Following Ref.~\cite{Diakonov:1997mm}, we use in this paper the decay operator
describing the emission of a $p-$wave pseudoscalar meson $\varphi$, which
has been obtained via the Goldberger-Treiman relation from the   collective
weak current \cite{Kim:2017khv},
\begin{align}
\mathcal{O}_{\varphi}&=\frac{1}{2F_{\varphi}}\left[  -\tilde{a}_{1}%
D_{\varphi\,i}^{(8)}-a_{2}\,d_{ibc}D_{\varphi\,b}^{(8)}\hat{J}_{c}-a_{3}%
\frac{1}{\sqrt{3}}D_{\varphi\,8}^{(8)}\hat{J}_{i}\right]  \,p_{i} .  \notag \\
& 
\label{eq:Oai}%
\end{align}
Constants $a_{1,2,3}$ that enter Eq.~(\ref{eq:Oai})
have
been extracted from the semileptonic decays of the baryon octet in
Ref.~\cite{Yang:2015era}:
\begin{equation}
a_{1}\simeq -3.509\,, \;a_{2} \simeq 3.437\, , \; a_{3}   \simeq 0.604\, .
\label{eq:a123}%
\end{equation}
However, due to the fact that $a_1$ scales as $N_{\rm val}$, it has been shown in
Ref.~\cite{Kim:2017khv} that in the heavy quark sector $a_1$ has to be replaced by
\begin{equation}
a_1 \rightarrow \tilde{a}_{1} =-2.1596. \label{eq:a1scaled}%
\end{equation}
With this replacement, all decays of charm and bottom sextet 
and of two exotic $\Omega_c$'s
have been successfully 
described by the present model \cite{Kim:2017khv}.
For the decay constants $F_{\varphi}$, we have adopted the convention in which
$F_{\pi}=93$~MeV and $F_{K}=F_{\eta}=1.2\,F_{\pi}=112$ MeV.

We are considering decays $B_{1}\rightarrow B_{2}+\varphi$, where $M_{1,2}$
denote masses of the initial and final baryons respectively, and $p_{i}$ is the
c.m. momentum of the outgoing meson of mass $m$~\cite{Kim:2017khv,Diakonov:1997mm}:%
\begin{equation}
\left\vert \vec{p}\, \right\vert =p=\frac{\sqrt{(M_{1}^{2}%
-(M_{2}+m)^{2})(M_{1}^{2}-(M_{2}-m)^{2})}}{2M_{1}}%
\end{equation}

The decay width is related to the matrix element of $\mathcal{O}_{\varphi}$
squared, summed over the final isospin (but not spin) and averaged over the
initial spin and isospin denoted as $\overline{\left[  \ldots\right]  ^{2}}$;
see the Appendix of Ref.~\cite{Diakonov:1997mm} 
and Erratum of Ref.~\cite{Kim:2017khv}
for the details of the
corresponding calculations,
\begin{equation}
\Gamma_{B_{1}\rightarrow B_{2}+\varphi}=\frac{1}{2\pi}\overline{\left\langle
B_{2}\left\vert \mathcal{O}_{\varphi}\right\vert B_{1}\right\rangle ^{2}%
}\,\frac{M_{2}}{M_{1}}p.
\end{equation}
Here factor $M_{2}/M_{1}$, used already in Ref.~\cite{Kim:2017khv}, is the same as in heavy baryon chiral perturbation
theory (HBChPT); see \emph{e.g.} Ref.~\cite{Cheng:2006dk}. 

\renewcommand{\arraystretch}{1.2}
\begin{table}[h!]
\parbox{.45\linewidth}{
\centering
\begin{tabular}{|c|c|c|}
\hline
$\overline{\boldsymbol{15}}_{J=1}$ & $\boldsymbol{6}_{J=1}$ &allowed \\\hline
$\Omega^c$ &$ \Omega^c+\pi$ & yes\\
& $\Xi^c+\overline{K}$ & ~yes$^*$ \\\hline
$\Xi^c_{3/2,1/2}$
& $\Xi^c+\pi$ & yes\\
&$ \Sigma^c+\overline{K}$ &~yes$^*$  \\\hline
$\Xi^c_{1/2}$ & $\Omega^c+{K}$ & no\\
& $\Xi^c+\eta$ & no\\ \hline
$\Sigma^c,\, \Lambda^c $& $\Xi^c+K$ & no\\
&$\Sigma^c+\pi$ & yes\\
\hline
$\Sigma^c$& $\Sigma^c +\eta$ & no\\ \hline
$N^c$ & $\Sigma^c+{K}$ & no\\
\hline
\end{tabular}
\caption{Decays of $\overline{\boldsymbol{15}}_{J=1}$ to $\boldsymbol{6}_{J=1}$.
($^*${only $s=3/2 \rightarrow s=1/2$}).}
\label{tab:decs15bar1a}} \hfill\parbox{.45\linewidth}{
\centering
\begin{tabular}{|c|c|c|}
\hline
$\overline{\boldsymbol{15}}_{J=1} $&$ {\overline{\boldsymbol{3}}}_{J=0}$ & {allowed}\\\hline
$\Omega^c$ & $\Xi^c+\overline{K} $& yes\\\hline
$\Xi^c_{3/2,1/2}$ & $\Xi^c+\pi$ & yes\\
\hline
$\Xi^c_{1/2}$ & $\Xi^c+\eta$ &~no$^\dagger$ \\
& $\Lambda^c+\overline{K} $& yes\\\hline
$\Sigma^c $& $\Xi^c+{K}$ &no \\
& $\Lambda^c+\pi $& yes\\
\hline
$\Lambda^c$ & $\Lambda^c+\eta $&~yes$^*$ \\
& $\Xi^c+{K}$ & no\\
\hline
$N^c $& $\Lambda^c+{K}$ &no \\\hline
\end{tabular}
\caption{Decays of $\overline{\boldsymbol{15}}_{J=1} $ to $ {\overline{\boldsymbol{3}}}_{J=0}$.
($^*${only $s=3/2$},~ $^\dagger$ $s=3/2$ at the threshold).}
\label{tab:decs15bar1b}}\end{table}

Because operator $\mathcal{O}_{\varphi}$ does not depend on the heavy quark
spin, it is only the soliton that decays by emitting a pseudoscalar meson.
Heavy quark acts as a spectator of the decaying soliton. Since the decay
occurs in the $p-$wave, the final soliton spin has to couple with meson angular
momentum $l=1$ to the spin of the initial state soliton. Decays with heavy
quark spin flip are suppressed by $1/m_{Q}$ and are not considered here.
Mesons are in the SU(3) octet and therefore the following decays are possible:%
\begin{align}
\overline{\boldsymbol{15}}_{J=1}  &  \rightarrow\boldsymbol{6}_{J=1}%
,\overline{\,\boldsymbol{3}}_{J=0},\nonumber\\
\overline{\boldsymbol{15}}_{J=0}  &  \rightarrow\overline{\boldsymbol{15}%
}_{J=1},\boldsymbol{6}_{J=1}.
\label{eq:despatterns}
\end{align}
Direct decays of $\overline{\boldsymbol{15}}_{J=0}$ to the ground state
antitriplet are suppressed. In Tables~\ref{tab:decs15bar1a} --
\ref{tab:decs15bar0b}, we list all decays that are allowed by the quantum numbers.
Taking into account mass estimates from the previous section,
we find that some of these decays are excluded by the energy conservation (column 3).
Furthermore, in Tab.~\ref{tab:decs15bartoD}, we list all possible decays of exotic 
$\overline{\boldsymbol{15}}$ to the ground state baryons and heavy mesons
(calculation of which is beyond the scope of the present paper).

\begin{table}[ptb]
\centering
\begin{tabular}{|c|c|c|}
\hline
$\overline{\boldsymbol{15}}_{J=0}$ & $\overline{\boldsymbol{15}}_{J=1}$ &allowed \\\hline
$\Omega^c$ &$ \Omega^c+\pi$ & yes\\
&$ \Omega^c+\eta$ &no \\
& $\Xi^c_{3/2,1/2}+\overline{K}$ &no \\\hline
$\Xi^c_{3/2,1/2}$ & $\Omega^c+{K}$ &no \\
& $\Xi^c_{3/2,1/2}+\pi$ &~yes$^*$ \\
&$ \Sigma^c+\overline{K}$ & no \\\hline
$\Xi^c_{1/2} $ & $\Xi^c_{1/2}+\eta$ & no\\
& $\Lambda^c+\overline{K}$& no \\\hline
$\Xi^c_{3/2} $ & $\Xi^c_{3/2}+\eta$ & no\\ \hline
$\Sigma^c $& $\Xi^c_{3/2,1/2}+K$ & no\\
& $\Lambda^c+\pi$ & yes\\
&$\Sigma^c+\pi$ & yes\\
&$\Sigma^c+\eta$ & no \\
& $N^c+\overline{K} $&no \\\hline
$\Lambda^c$ &$ \Xi^c_{1/2}+K$ &no \\
& $\Lambda^c + \eta $ &no \\
& $\Sigma^c+\pi$ & ~yes$^{\dagger}$\\
&$ N^c+\overline{K} $&no \\\hline
$N^c$ & $\Sigma^c+{K}$ &no \\
& $\Lambda^c+{K}$ & no\\
&$N^c+\pi$&~yes$^\#$ \\
&$N^c+\eta$& no\\
\hline
\end{tabular}
\caption{Decays of $\overline{\boldsymbol{15}}_{J=0}$ to $\overline{\boldsymbol{15}}_{J=1}$.
($^*~\Xi_{1/2}\rightarrow \Xi_{3/2}(s=3/2)$ below the threshold,~ $^{\dagger}~s=1/2
\rightarrow s=3/2$~below the threshold,~
$^{\#}$~only $s=3/2
\rightarrow s=1/2$~allowed).}
\label{tab:decs15bar0a}
\end{table}
\begin{table}[h!]
\parbox{.45\linewidth}{
\centering
\begin{tabular}{|c|c|c|}
\hline
$\overline{\boldsymbol{15}}_{J=0}$ & $\boldsymbol{6}_{J=1}$ &allowed \\\hline
$\Omega^c$ &$ \Omega^c+\pi$ & yes\\
& $\Xi^c+\overline{K}$ &yes \\\hline
$\Xi^c_{3/2,1/2}$
& $\Xi^c+\pi$ & yes\\
&$ \Sigma^c+\overline{K}$ & yes\\\hline
$\Xi^c_{1/2}$  & $\Omega^c+{K}$ &no \\
& $\Xi^c + \eta$ & ~no$^*$\\ \hline
$\Sigma^c, \Lambda $& $\Xi+K$ &no \\
&$\Sigma^c+\pi$ & yes\\ \hline
$\Sigma^c$ & $\Sigma^c+\eta$ &~yes$^\dagger$ \\ \hline
$N^c$ & $\Sigma^c+{K}$ & no \\
\hline
\end{tabular}
\caption{Decays of $\overline{\boldsymbol{15}}_{J=0}$ to $\boldsymbol{6}_{J=1}$.
($^*$decay to $s=1/2$ at the threshold,
$^\dagger$only to $s=1/2$).}
\label{tab:decs15bar0b}} \hfill
\parbox{.45\linewidth}{
\centering
\begin{tabular}[c]{|c|c|c|c|}
\hline
$\overline{\boldsymbol{15}}$ & final & $J=1$ & $J=0 $\\\hline
$\Omega^{c}$ & $\Xi+D$ & no & yes\\\hline
$\Xi^{c}_{3/2}$ & $\Sigma+D$ & no & yes\\\hline
$\Xi^{c}_{1/2}$ & $\Xi+D_{s}$ & no & no\\
& $\Lambda+D $ & no & yes\\
& $\Sigma+D$ & no & yes\\\hline
$\Sigma^{c}$ & $\Sigma+D_{s} $ & no & no\\
& $N+D$ & yes$^*$ & yes\\\hline
$\Lambda^{c}$ & $\Lambda+D_{s}$ & no & no\\
& $N+D$ & yes$^*$ & yes\\\hline
$N^{c}$ & $N+D_{s}$ & no & no\\\hline
\end{tabular}
\caption{Decays of $\overline{\boldsymbol{15}}_{J=1,0} $ to the ground state
octet baryons and $D$ mesons ($^*$only $s=3/2$). }%
\label{tab:decs15bartoD}}%
\end{table}

Already at this point we can draw interesting conclusions. For $\overline{\boldsymbol{15}}_{J=1} $,
we have ten allowed decays of the type (\ref{eq:despatterns}) and for
$\overline{\boldsymbol{15}}_{J=0} $ twelve. Interestingly, the lightest members of both multiplets,
namely $N_{\overline{\boldsymbol{15}}}$ are stable against two body strong interactions. Except for 
$N_{\overline{\boldsymbol{15}}}$, all members of $J=0$ multiplet have open channels to the decays
to heavy mesons.

After averaging over the initial spin and isospin and summing over the final isospin and over final spin third component
$m_2$, we arrive at the following expressions for the decay widths:
\begin{align}
\Gamma_{B_{1}\rightarrow B_{2}+\varphi} 
=&  \frac{p^{3}}{24\pi F_{\varphi}^{2}}\frac{M_{2}}{M_{1}} \,
 \frac{\dim\mathcal{R}_{2}}{\dim\overline{\boldsymbol{15}}} 
 \; \gamma_{J_1\rightarrow J_2}(s_{1}\rightarrow s_{2})
  \notag \\
\times &\left( \sum_{\mu}  {\mathcal {G}}_{\overline{\boldsymbol{15}}_{J_1} \rightarrow \mathcal{R}_{2}}^{(\mu)}
\left[
\begin{array}
[c]{cc}%
8 & \mathcal{R}_{2}\\
\varphi & B_{2}%
\end{array}
\right\vert \left.
\begin{array}
[c]{c}%
\overline{\boldsymbol{15}}_{\mu}\\
B_{1}%
\end{array}
\right] \right)^{2} \, .
\label{eq:Gammas}
\end{align}
Here the square bracket stands for the pertinent SU(3) isoscalar factor coupling
meson and baryon in a final state to the baryon in the initial state and
$ {\mathcal {G}}^{(\mu)}_{\overline{\boldsymbol{15}}_{J_1}\rightarrow \mathcal{R}_{2}}$ is the decay coupling 
(see below).\footnote{Recall that the soliton in SU(3) $\overline{\boldsymbol{15}}$ can be quantized
as spin $J_1=1$ or $J_1=0$.}
The sum over $\mu$ is relevant only for $\mathcal{R}_{2}=\overline{\boldsymbol{15}}$ in the final state.
Here, we adopt the de Swart conventions for the SU(3) phase factors \cite{deSwart:1963pdg} and label 
the representations as in the numerical code of Ref.~\cite{Kaeding:1995re}.
Factors $\gamma$ take care of the spin dependence for the soliton angular momenta  $J=1$ or 0
(see Erratum in Ref.~\cite{Kim:2017khv}):%
\begin{align}
\gamma_{1 \rightarrow 1}(1/2 \rightarrow1/2)=2/3,
&\qquad
\gamma_{1\rightarrow 1}(1/2\rightarrow3/2)=1/3, 
\notag \\
\gamma_{1\rightarrow 1}(3/2 \rightarrow1/2)=1/6,
&\qquad
\gamma_{1\rightarrow 1}(3/2\rightarrow3/2)=5/6, 
\notag \\
\gamma_{0 \rightarrow 1}(1/2\rightarrow1/2)=1/3, 
&\qquad
\gamma_{0 \rightarrow 1}(1/2\rightarrow3/2)=2/3, 
\notag \\
\gamma_{1 \rightarrow 0}(1/2\rightarrow 1/2)=1,~~\,
&\qquad
\gamma_{1 \rightarrow 0}(3/2\rightarrow 1/2)=1.
\end{align}
Note that
\begin{equation}
\sum_{s_2} \gamma_{J_1 \rightarrow J_2}(s_1\rightarrow s_2)=1\, .
\end{equation}

Finally, the decay constants read\footnote{In the present definition of the decay constants we have included
a pertinent SU(3) {\em spin} isoscalar factor, which has not been included in definitions of Ref.~\cite{Kim:2017khv}.}
\begin{align}
{\mathcal {G}}_{{\overline{\boldsymbol{15}}_1}\rightarrow {\overline{\boldsymbol{3}}}_0} &=
\sqrt{\frac{1}{2}}
\left(  -\tilde{a}_{1}-\frac
{1}{2}a_{2}\right)
=0.312 \, ,
 \notag \\
{\mathcal {G}}_{\overline{\boldsymbol{15}}_1\rightarrow {\boldsymbol{6}}_1} &=
-\sqrt{\frac{1}{3}}
\left(  -\tilde{a}_{1}-\frac
{1}{2}a_{2}-a_{3}\right)
=0.094\, ,
 \notag \\
{\mathcal {G}}_{\overline{\boldsymbol{15}}_0\rightarrow {\boldsymbol{6}}_1} &=
\frac{1}{2}
\left(  -\tilde{a}_{1}-\frac
{3}{2}a_{2}\right)
=-1.498\, ,
\notag \\
{\mathcal {G}}^{(\mu=1)}_{{\overline{\boldsymbol{15}}_0}\rightarrow {\overline{\boldsymbol{15}}}_1} &=
\sqrt{\frac{1}{366}}\left(
-\tilde{a}_{1}+\frac{41}{2}a_{2}\right) 
=3.796 \, ,
\notag \\
{\mathcal {G}}^{(\mu=2)}_{{\overline{\boldsymbol{15}}_0}\rightarrow {\overline{\boldsymbol{15}}}_1} &=
-\sqrt{\frac{81}{122}}\left(
-\tilde{a}_{1}+\frac{1}{6}a_{2}\right) 
=- 2.226 \, .
\label{eq:decayconsts}
\end{align}
To compute the numerical values, we have used Eqs.~(\ref{eq:a123}) and (\ref{eq:a1scaled}).
We see from Eqs.~(\ref{eq:decayconsts}) that decay constants of $(\overline{\boldsymbol{15}},J=1)$ are very small;
in fact they vanish in the large $N_c$ limit \cite{Praszalowicz:2018upb}. On the contrary, decay constants of 
$(\overline{\boldsymbol{15}},J=0)$ are almost an order of magnitude larger, so we expect the corresponding decay widths
to be large (furthermore, the phase space factor $p^3$ will be larger than in the $J=1$ case).

\subsection{Numerical estimates}

Numerical estimates of the decay widths, assuming central values for baryon masses from Table~\ref{tab:mpred},
are listed in Tables \ref{tab:decs15bar1} and  \ref{tab:decs15bar0}. One should note that these widths are pure 
predictions based on the light sector va{\-}lu{\-}es of the decay parameters, except for rescaling (\ref{eq:a1scaled}).

Decay widths of the $\Omega^c$ states from $\overline{\boldsymbol{15}}_{J=1}$, $\Gamma=0.43$~MeV and  0.98~MeV
have been already computed 
in Ref.~\cite{Kim:2017khv} and agree within uncertainties with experimental widths equal to $0.8\pm 0.2\pm 0.1$~MeV
and $1.1\pm 0.8 \pm 0.4$~MeV, respectively~\cite{LHCb:2017uwr}. Our present results also agree with initial estimates
of the decay widths of two exotic $\Xi^c_{3/2}$ states given in  Ref.~\cite{Kim:2017khv} (note that here we have slightly different
masses). 

We see from Table~\ref{tab:decs15bar1} that all states in  $\overline{\boldsymbol{15}}_{J=1}$, have very small
widths, in most cases not exceeding 1~MeV. For almost all states in $\overline{\boldsymbol{15}}_{J=1}$, decay channels 
to light baryons and heavy mesons are closed (except for $\Sigma^c$ and $\Lambda^c$ of spin $s=3/2$, which are at the threshold).
We therefore conclude that exotic charm pentaquarks from $\overline{\boldsymbol{15}}_{J=1}$ can be found only in dedicated searches 
in high resolution experiments.
One should also observe that, as already shown in Tables~\ref{tab:decs15bar1a} and \ref{tab:decs15bar1b}, that the lightest
member of $\overline{\boldsymbol{15}}_{J=1}$, namely the nucleonlike pentaquark, is stable with respect to two body strong
decays.

\linespread{1.25}
\begin{table*}[t]
\centering
\begin{tabular}{|c|c|c|c|c|}

\hline
\multicolumn{2}{|c|}{decay}&\multicolumn{3}{c|}{$\Gamma$~[MeV]}\\
\hline
$B_1$  & $B_2+\varphi$&  $s_2=\frac{1}{2}$ & $s_2=\frac{3}{2}$ & $\Sigma_{s_2}$\\ 
\hline
\multirow{2}{*}{$\Omega^c({\boldsymbol{\overline{15}}}_{1}^{1/2})$} &$\Xi^c({\overline{\boldsymbol{3}}}_{0})+\overline{K}$& 0.349& -- -- --&0.349   \\ 
 & $\Omega^c({\boldsymbol{6}}_{1})+\pi$& 0.062& 0.015& 0.077  \\ 
 \cline{2-5} 
&\multicolumn{3}{l|}{~~~total}& 0.425 \\ \hline
 &$\Xi^c({\overline{\boldsymbol{3}}}_{0})+\overline{K}$& 0.875& -- -- --&0.875   \\ 
$\Omega^c({\boldsymbol{\overline{15}}}_{1}^{3/2})$ & $\Omega^c({\boldsymbol{6}}_{1})+\pi$& 0.027& 0.077& 0.104  \\ 
&$\Xi^c({{\boldsymbol{6}}}_{1})+\overline{K}$&0.002&-- -- --&0.002 \\
 \cline{2-5} 
&\multicolumn{3}{l|}{~~~total}& 0.981 \\ \hline
\multirow{2}{*}{$\Xi^c_{3/2}({\overline{\boldsymbol{15}}}_{1}^{1/2})$}& $\Xi^c({\overline{\boldsymbol{3}}}_{0})+\pi$&1.636 -- 1.830 & -- -- --&1.636 -- 1.830 \\ 
  & $\Xi^c({\boldsymbol{6}}_{1})+\pi$&0.029 --0.035 &0.007 -- 0.009 & 0.036 -- 0.043  \\  
 \cline{2-5} 
&\multicolumn{3}{l|}{~~~total}& 1.672 -- 1.874 \\ \hline
& 
 $\Xi^c({\overline{\boldsymbol{3}}}_{0})+\pi$&2.447 -- 2.687 & -- -- --&2.447 -- 2.687\\ 
$\Xi^c_{3/2}({\overline{\boldsymbol{15}}}_{1}^{3/2})$ & $\Xi^c({\boldsymbol{6}}_{1})+\pi$&0.013 -- 0.015 &0.037 -- 0.044 & 0.050 -- 0.058  \\
& $\Sigma^c({\boldsymbol{6}}_{1})+\overline{K}$ & 0.003 -- 0.005 & $\approx$ 0 & 0.003 -- 0.005 \\ 
 \cline{2-5} &\multicolumn{3}{l|}{~~~total}& 2.497 -- 2.751 \\ \hline
 &   $\Xi^c({\boldsymbol{3}}_{0})+\pi$&0.092 -- 0.105 &-- -- --&0.092 -- 0.105\\ 
$\Xi^c_{1/2}({\overline{\boldsymbol{15}}}_{1}^{1/2})$&$\Lambda^c({\overline{\boldsymbol{3}}}_{0})+\overline{K}$&0.239 -- 0.299 & -- -- --&0.239 -- 0.299 \\ 
 & $\Xi^c({\overline{\boldsymbol{6}}}_{1})+\pi$&0.039 -- 0.048 & 0.009 -- 0.011 &0.048 -- 0.059\\  
 \cline{2-5} 
&\multicolumn{3}{l|}{~~~total}& 0.379 -- 0.463\\ \hline
  & $\Xi^c({\boldsymbol{3}}_{0})+\pi$&0.140 -- 0.156 &-- -- --& 0.140 -- 0.156\\ 
\multirow{2}{*}{$\Xi^c_{1/2}({\overline{\boldsymbol{15}}}_{1}^{3/2})$}&  $\Lambda^c({\overline{\boldsymbol{3}}}_{0})+\overline{K}$&0.468 -- 0.546 & -- -- --&0.468 -- 0.546\\ 
 & $\Xi^c({\overline{\boldsymbol{6}}}_{1})+\pi$&0.018 -- 0.021 & 0.049 -- 0.060& 0.068 -- 0.081  \\ 
 & $\Sigma^c({\boldsymbol{6}}_{1})+\overline{K}$ &  $\approx$ 0 & -- -- -- & $\approx$ 0 \\ 
 \cline{2-5} 
 &\multicolumn{3}{l|}{~~~total}& 0.676 -- 0.783 \\ \hline
 \multirow{2}{*}{$\Sigma^c({\overline{\boldsymbol{15}}}_{1}^{1/2})$}&  $\Lambda^c({\overline{\boldsymbol{3}}}_{0})+\pi$&1.073 -- 1.165 &-- -- --&1.073 -- 1.165 \\ 
  & $\Sigma^c({\boldsymbol{6}}_{1})+\pi$&0.027 -- 0.031 &0.006 -- 0.008 & 0.033 -- 0.039 \\ 
 \cline{2-5} 
 &\multicolumn{3}{l|}{~~~total}& 1.107 -- 1.203 \\ \hline
  \multirow{2}{*}{$\Sigma^c({\overline{\boldsymbol{15}}}_{1}^{3/2})$} &  $\Lambda^c({\overline{\boldsymbol{3}}}_{0})+\pi$&1.525 -- 1.635&--  -- --&1.525 -- 1.635  \\ 
 & $\Sigma^c({\boldsymbol{6}}_{1})+\pi$&0.012 -- 0.013 &0.035 -- 0.040 &0.047 -- 0.054\\ 

 \cline{2-5} 
 &\multicolumn{3}{l|}{~~~total}& 1.572 -- 1.688\\ \hline
 $\Lambda^c({\overline{\boldsymbol{15}}}_{1}^{1/2})$ & $\Sigma^c({\boldsymbol{6}}_{1})+\pi$&0.016 -- 0.023 &0.004 --0.006 & 0.019 -- 0.030\\ 
 \cline{2-5} 
 &\multicolumn{3}{l|}{~~~total}&0.019 -- 0.030 \\ \hline
  \multirow{2}{*}{$\Lambda^c({\overline{\boldsymbol{15}}}_{1}^{3/2})$}  & $\Lambda^c({\overline{\boldsymbol{3}}}_{0}) + \eta$ & 0.006 -- 0.108  & -- -- -- & 0.006 --0.108 \\ 
& $\Sigma^c({\boldsymbol{6}}_{1})+\pi$&0.008 -- 0.010 &0.021 -- 0.031 & 0.028 -- 0.041 \\ 
 \cline{2-5}
  &\multicolumn{3}{l|}{~~~total}&0.034 -- 0.149\\ \hline
\end{tabular}
\caption{Decay widths of exotic pentaquarks in the SU(3) representation $\overline{\boldsymbol{15}}$ and $J=1$.
Uncertainties correspond to the
mass ranges from Table \ref{tab:mpred} (two 
$\Omega^c$ states are used as the input; therefore, their masses and decays widths are not subject to
such uncertainties).
\label{tab:decs15bar1}}
\end{table*}
\linespread{1}

\linespread{1.25}
\begin{table*}[t]
\centering
\begin{tabular}{|c|c|c|c|c|}
\hline
\multicolumn{2}{|c|}{decay}&\multicolumn{3}{c|}{$\Gamma$~[MeV]}\\
\hline
$B_1$  & $B_2+\varphi$&  $s_2=\frac{1}{2}$ & $s_2=\frac{3}{2}$ & $\Sigma_{s_2}$\\ 
\hline
 & $\Omega^c({\boldsymbol{6}}_{1})+\pi$& 34.18 -- 41.02 & 47.93 -- 59.38 &82.10 -- 100.39   \\ 
 $\Omega^c({\overline{\boldsymbol{15}}}_{0})$& $\Xi_{1/2}({{\boldsymbol{6}}}_{1})+\overline{K}$& 7.14 -- 9.53 & 7.5 -- 11.31& 14.63 -- 20.84  \\ 
 &$\Omega^c({\overline{\boldsymbol{15}}}_{1})+\pi$& 1.49 -- 2.97& 0.23 -- 1.48& 1.72 -- 4.46   \\ 
 \cline{2-5} 
&\multicolumn{3}{l|}{~~~total}& 98.46 -- 125.68 \\ \hline
 & $\Xi^c({\boldsymbol{6}}_{1})+\pi$&17.52 -- 20.48 &25.04 -- 30.06 & 42.56 --50.55  \\ 
 \multirow{2}{*}{$\Xi^c_{3/2}({\overline{\boldsymbol{15}}}_{0})$}& 
 $\Sigma^c({\boldsymbol{6}}_{1})+\overline{K}$&15.50 -- 19.7 & 17.99 -- 24.90& 33.49 -- 44.60  \\ 
 &$\Xi^c_{3/2}({\overline{\boldsymbol{15}}}_{1})+\pi$&22.83 -- 56.04 & 2.65 -- 31.53 & 25.48 -- 87.57 \\ 
 &$\Xi^c_{1/2}({\overline{\boldsymbol{15}}}_{1})+\pi$& 2.66 -- 6.11 & 0.67 -- 4.04 & 3.33 -- 10.16 \\ 
 \cline{2-5} 
&\multicolumn{3}{l|}{~~~total}& 104.86 -- 192.89\\ \hline
 & $\Xi^c({\boldsymbol{6}}_{1})+\pi$&23.83 -- 25.01 &33.21 -- 35.19 & 57.04 -- 60.21  \\ 
 \multirow{2}{*}{$\Xi^c_{1/2}({\overline{\boldsymbol{15}}}_{0})$}& 
 $\Sigma^c({\boldsymbol{6}}_{1})+\overline{K}$&0.77 -- 0.84 & 0.81 -- 0.91 & 1.58 -- 1.75  \\  
 &$\Xi^c_{3/2}({\overline{\boldsymbol{15}}}_{1})+\pi$&1.93 -- 3.95& 0 -- 0.42 & 1.93 -- 4.37   \\ 
 &$\Xi^c_{1/2}({\overline{\boldsymbol{15}}}_{1})+\pi$& 0.09  -- 0.17  & 0 -- 0.05 & 0.09 -- 0.22 \\ 
 \cline{2-5} 
&\multicolumn{3}{l|}{~~~total}& 60.64 -- 66.55 \\ \hline
 & $\Sigma^c({\boldsymbol{6}}_{1})+\pi$& 13.38 -- 17.08& 18.83 -- 25.06& 32.21 -- 42.14 \\ 
 \multirow{2}{*}{ $\Sigma^c({\overline{\boldsymbol{15}}}_{0})$} & $\Sigma^c({\boldsymbol{6}}_{1})+\eta$& 0 -- 0.95 & -- -- --& 0 -- 0.95 \\ 
& $\Sigma^c({\overline{\boldsymbol{15}}}_{1})+\pi$&7.31 -- 34.90 &0 -- 12.55 &7.31 -- 47.45 \\ 
 &$\Lambda^c({\overline{\boldsymbol{15}}}_{1})+\pi$& 0.95 -- 6.95& 0 -- 3.74& 0.95 -- 10.68\\ 
 \cline{2-5} 
&\multicolumn{3}{l|}{~~~total}& 40.46 -- 101.23\\ \hline
 \multirow{2}{*}{$\Lambda^c({\overline{\boldsymbol{15}}}_{0})$} & $\Sigma^c({\boldsymbol{6}}_{1})+\pi$&~9.42 -- 10.46 &13.10 -- 14.82 & 22.53 -- 25.27  \\ 
& 
 $\Sigma^c({\overline{\boldsymbol{15}}}_{1})+\pi$&1.84 -- 6.21&  -- -- -- & 1.84 -- 6.21 \\ 
 \cline{2-5} 
&\multicolumn{3}{l|}{~~~total}& 28.74 -- 31.48\\ \hline
 $N^c({\overline{\boldsymbol{15}}}_{0})$& $N^c({\overline{\boldsymbol{15}}}_{1})+\pi$&0 -- 34.47 & 0 -- 9.98& 0 -- 44.45\\ \cline{2-5} 
 &\multicolumn{3}{l|}{~~~total}& 0 -- 44.45\\ \hline
\end{tabular}
\caption{Decay widths of exotic pentaquarks in the SU(3) representation $\overline{\boldsymbol{15}}$ and $J=0$.
Uncertainties correspond to the
mass ranges from Table \ref{tab:mpred}.
\label{tab:decs15bar0}}
\end{table*}

The situation is completely different in the case of $\overline{\boldsymbol{15}}_{J=0}$ listed in Table~\ref{tab:decs15bar0}.
Here all decay widths 
are within 30~--~140~MeV range. The only exception is again the lightest nucleonlike pentaquark, which however,
can decay only to the $N^c$ state in $\overline{\boldsymbol{15}}_{J=1}$, which is semistable. Furthermore,
all states in $\overline{\boldsymbol{15}}_{J=0}$ (except for $N^c$) have at least one open channel to the decays
to light baryons and heavy mesons. We are not able to compute these widths within the present approach.
However, since the available phase space is comparable to the decays listed in Tab.~\ref{tab:decs15bar0},
we may expect that the total decay widths will double with respect to the estimates given in Tab.~\ref{tab:decs15bar0}.

One should also note, that all decays of $\overline{\boldsymbol{15}}_{J=0}$ lead to either $\boldsymbol{6}$
or $\overline{\boldsymbol{15}}_{J=1}$, which decay further to $\boldsymbol{6}$ and $\overline{\boldsymbol{3}}$.

We conclude therefore, that pentaquarks from $\overline{\boldsymbol{15}}_{J=0}$ multiplet are very wide and may be
interpreted as a background, rather than as a signal. Therefore, they could have been missed in general purpose experiments.

\section{Summary and conclusions}
\label{sec:conclusions}

In the present paper we have studied the consequences of possible existence of heavy pentaquark
SU(3) multiplets. Charmed pentaquarks
have been evoked to explain small widths of two excited $\Omega^c$ states \cite{Kim:2017jpx} announced
in 2017 by the LHCb Collaboration at CERN \cite{LHCb:2017uwr}. Such interpretation requires, however,
the existence of many other exotic and cryptoexotic charm baryons that have not been observed experimentally.

For the present study we have employed the $\chi$QSM estimating its parameters from the heavy baryon spectra.
Therefore strictly speaking, we have not tested the dynamics of the model, but rather the underlying {\em hedgehog} SU(3) symmetry.
Such symmetry leads to the sum rules (\ref{eq:GMO}) analogous to the Gell-Mann--Okubo mass relations
\cite{Gell-Mann:1962yej,Okubo:1961jc} and to one Guadagnini-type \cite{Guadagnini:1983uv} relation (\ref{eq:Gaud}).

We presented numerical support for the model mass formulas (\ref{eq:15barav}), (\ref{eq:Delta15bar}) and (\ref{eq:M3barM6mass}).
Next, we extracted model parameters from the heavy baryon spectra alone, and from the positivity of splitting
parameters $\alpha,~\beta$, and $\gamma$ (\ref{eq:Hsb}). We obtained mass ranges of the charm pentaquarks
with uncertainties of the order $\sim50$~MeV. Of course this  is a conservative estimate, as the model itself is
to large extent semiquantitave.

Finally, we computed the decay widths. Here, predictions for known experimentally ground state charm baryons 
as well as for two exotic $\Omega^c$
states, are very accurate \cite{Kim:2017khv}. We therefore have confidence in our predictions for the remaining 
exotic states.

We have found that pentaquarks belonging to the $\overline{\boldsymbol{15}}_{J=1}$ SU(3) multiplet
are very narrow having widths of the order of $\sim 1$~MeV, while the remaining states from the
$\overline{\boldsymbol{15}}_{J=0}$ SU(3) multiplet are wide, in most cases of the order of $\sim 100$~MeV or more.
Moreover, all these decays lead to the unstable resonances; therefore, the identification of exotica requires
 dedicated experiments. Multipurpose searches could easily miss narrow or wide exotic states.

\section*{Acknowledgments}
This work has been supported by the Polish National Science Centre Grants No. 2017/27/B/ST2/01314 (MK and MP)
and No. 2018/31/B/ST2/01022 (MP).
MP thanks also the Institute for Nuclear Theory at the University of Washington for its kind hospitality and stimulating research environment. MP research was supported in part by the INT's U.S. Department of Energy Grant No. DE-FG02- 00ER41132.



\begin{thebibliography}{99}                     


\bibitem{Cheng:2021qpd}
H.~Y.~Cheng,
Chin. J. Phys. \textbf{78}, 324-362 (2022)
[arXiv:2109.01216 [hep-ph]].


\bibitem {PDG}P.A. Zyla et al. (Particle Data Group), Prog. Theor. Exp. Phys.
\textbf{083C01} (2020) and 2021 update.

\bibitem {Yang:2016qdz}G.~S.~Yang, H.-Ch.~Kim, M.~V.~Polyakov and
M.~Praszalowicz,
Phys.\ Rev.\ D \textbf{94}, 071502 (2016).

\bibitem {Kim:2017jpx}H.~C.~Kim, M.~V.~Polyakov and M.~Prasza\l {}owicz,
Phys. Rev. D \textbf{96} (2017) no.1, 014009 doi:10.1103/PhysRevD.96.014009
[arXiv:1704.04082 [hep-ph]].

\bibitem{Kim:2017khv}
H.~C.~Kim, M.~V.~Polyakov, M.~Praszalowicz and G.~S.~Yang,
Phys. Rev. D \textbf{96}, no.9, 094021 (2017)
[erratum: Phys. Rev. D \textbf{97}, no.3, 039901 (2018)]
doi:10.1103/PhysRevD.96.094021
[arXiv:1709.04927 [hep-ph]].

\bibitem{Polyakov:2022eub}
M.~V.~Polyakov and M.~Praszalowicz,
Phys. Rev. D \textbf{105}, 094004 (2022)
doi:10.1103/PhysRevD.105.094004
[arXiv:2201.07293 [hep-ph]].

\bibitem{Praszalowicz:2003ik}
M.~Praszalowicz,
Phys. Lett. B \textbf{575}, 234-241 (2003)
doi:10.1016/j.physletb.2003.09.049
[arXiv:hep-ph/0308114 [hep-ph]].

\bibitem{Diakonov:1997mm}
D.~Diakonov, V.~Petrov and M.~V.~Polyakov,
Z. Phys. A \textbf{359}, 305-314 (1997)
doi:10.1007/s002180050406
[arXiv:hep-ph/9703373 [hep-ph]].

\bibitem {LHCb:2017uwr}R.~Aaij \textit{et al.} [LHCb],
Phys. Rev. Lett. \textbf{118}, no.18, 182001 (2017)
doi:10.1103/PhysRevLett.118.182001 [arXiv:1703.04639 [hep-ex]].


\bibitem {Belle:2017ext}J.~Yelton \textit{et al.} [Belle],
Phys. Rev. D \textbf{97}, no.5, 051102 (2018) doi:10.1103/PhysRevD.97.051102
[arXiv:1711.07927 [hep-ex]].


\bibitem{LHCb:2021ptx}
R.~Aaij \textit{et al.} [LHCb],
Phys. Rev. D \textbf{104}, no.9, 9 (2021)
doi:10.1103/PhysRevD.104.L091102
[arXiv:2107.03419 [hep-ex]].

\bibitem{Karliner:2017kfm}
M.~Karliner and J.~L.~Rosner,
Phys. Rev. D \textbf{95}, no.11, 114012 (2017)
doi:10.1103/PhysRevD.95.114012
[arXiv:1703.07774 [hep-ph]].

\bibitem{Wang:2017vnc}
W.~Wang and R.~L.~Zhu,
Phys. Rev. D \textbf{96}, no.1, 014024 (2017)
doi:10.1103/PhysRevD.96.014024
[arXiv:1704.00179 [hep-ph]].

\bibitem{Chen:2017gnu}
B.~Chen and X.~Liu,
Phys. Rev. D \textbf{96}, no.9, 094015 (2017)
doi:10.1103/PhysRevD.96.094015
[arXiv:1704.02583 [hep-ph]].

\bibitem{Santopinto:2018ljf}
E.~Santopinto, A.~Giachino, J.~Ferretti, H.~Garc\'\i{}a-Tecocoatzi, M.~A.~Bedolla, R.~Bijker and E.~Ortiz-Pacheco,
Eur. Phys. J. C \textbf{79}, no.12, 1012 (2019)
doi:10.1140/epjc/s10052-019-7527-4
[arXiv:1811.01799 [hep-ph]].

\bibitem{Jia:2020vek}
D.~Jia, J.~H.~Pan and C.~Q.~Pang,
Eur. Phys. J. C \textbf{81}, no.5, 434 (2021)
doi:10.1140/epjc/s10052-021-09205-6
[arXiv:2007.01545 [hep-ph]].

\bibitem{Wang:2017zjw}
Z.~G.~Wang,
Eur. Phys. J. C \textbf{77}, no.5, 325 (2017)
doi:10.1140/epjc/s10052-017-4895-5
[arXiv:1704.01854 [hep-ph]].

\bibitem{Padmanath:2017lng}
M.~Padmanath and N.~Mathur,
Phys. Rev. Lett. \textbf{119}, no.4, 042001 (2017)
doi:10.1103/PhysRevLett.119.042001
[arXiv:1704.00259 [hep-ph]].

\bibitem{Cheng:2021qpd}
H.~Y.~Cheng,
Chin. J. Phys. \textbf{78}, 324-362 (2022)
doi:10.1016/j.cjph.2022.06.021
[arXiv:2109.01216 [hep-ph]].

\bibitem{An:2017lwg}
C.~S.~An and H.~Chen,
Phys. Rev. D \textbf{96}, no.3, 034012 (2017)
doi:10.1103/PhysRevD.96.034012
[arXiv:1705.08571 [hep-ph]].

\bibitem{Yang:2017rpg}
G.~Yang and J.~Ping,
Phys. Rev. D \textbf{97}, no.3, 034023 (2018)
doi:10.1103/PhysRevD.97.034023
[arXiv:1703.08845 [hep-ph]].

\bibitem{Wang:2018alb}
Z.~G.~Wang and J.~X.~Zhang,
Eur. Phys. J. C \textbf{78}, no.6, 503 (2018)
doi:10.1140/epjc/s10052-018-5989-4
[arXiv:1804.06195 [hep-ph]].

\bibitem{Wang:2017smo}
C.~Wang, L.~L.~Liu, X.~W.~Kang, X.~H.~Guo and R.~W.~Wang,
Eur. Phys. J. C \textbf{78}, no.5, 407 (2018)
doi:10.1140/epjc/s10052-018-5874-1
[arXiv:1710.10850 [hep-ph]].

\bibitem{Kim:2019rcx}
J.~Y.~Kim and H.~C.~Kim,
PTEP \textbf{2020}, no.4, 043D03 (2020)
doi:10.1093/ptep/ptaa037
[arXiv:1909.00123 [hep-ph]].

\bibitem {Diakonov:1987ty} D.~Diakonov, V.~Y.~Petrov and P.~V.~Pobylitsa,
Nucl.\ Phys.\ B \textbf{306} (1988) 809.




\bibitem {Christov:1995vm} C.~V.~Christov, A.~Blotz, H.~C.~Kim, P.~Pobylitsa,
T.~Watabe, T.~Meissner, E.~Ruiz Arriola and K.~Goeke,
Prog.\ Part.\ Nucl.\ Phys.\ \textbf{37} (1996) 91.




\bibitem {Alkofer:1994ph} R.~Alkofer, H.~Reinhardt and H.~Weigel,
Phys.\ Rept.\ \textbf{265} (1996) 139.




\bibitem {Petrov:2016vvl} V.~Petrov,
Acta Phys.\ Polon.\ B \textbf{47} (2016) 59.

\bibitem {Witten:1979kh}E.~Witten,
Nucl.\ Phys.\ B \textbf{160} (1979) 57.
\bibitem{WittenCA}
E.~Witten,
Nucl.\ Phys.\ B
\textbf{223} (1983) 422, and
\textbf{223} (1983) 433.

\bibitem{Adkins:1983ya}
G.~S.~Adkins, C.~R.~Nappi and E.~Witten,
Nucl. Phys. B \textbf{228}, 552 (1983)
doi:10.1016/0550-3213(83)90559-X

\bibitem{Wess:1971yu}
J.~Wess and B.~Zumino,
Phys. Lett. B \textbf{37}, 95-97 (1971)
doi:10.1016/0370-2693(71)90582-X

\bibitem{Guadagnini:1983uv}
E.~Guadagnini,
Nucl. Phys. B \textbf{236}, 35-47 (1984)
doi:10.1016/0550-3213(84)90523-6

\bibitem{Mazur:1984yf}
P.~O.~Mazur, M.~A.~Nowak and M.~Praszalowicz,
Phys. Lett. B \textbf{147}, 137-140 (1984)
doi:10.1016/0370-2693(84)90608-7

\bibitem{Jain:1984gp}
S.~Jain and S.~R.~Wadia,
Nucl. Phys. B \textbf{258}, 713 (1985)
doi:10.1016/0550-3213(85)90632-7

\bibitem {Diakonov:2013qta}D.~Diakonov, V.~Petrov and A.~A.~Vladimirov,
Phys. Rev. D \textbf{88}, no.7, 074030 (2013) doi:10.1103/PhysRevD.88.074030
[arXiv:1308.0947 [hep-ph]].

\bibitem {Adkins:1984cf} G.~S.~Adkins and C.~R.~Nappi,
Nucl.\ Phys.\ B \textbf{249}, 507 (1985).

\bibitem{Blotz:1992pw}
A.~Blotz, D.~Diakonov, K.~Goeke, N.~W.~Park, V.~Petrov and P.~V.~Pobylitsa,
Nucl. Phys. A \textbf{555}, 765-792 (1993)
doi:10.1016/0375-9474(93)90505-R

\bibitem{Gell-Mann:1962yej}
M.~Gell-Mann,
Phys. Rev. \textbf{125}, 1067-1084 (1962)
doi:10.1103/PhysRev.125.1067

\bibitem{Okubo:1961jc}
S.~Okubo,
Prog. Theor. Phys. \textbf{27}, 949-966 (1962)
doi:10.1143/PTP.27.949

\bibitem{Praszalowicz:2022sqx}
M.~Praszalowicz,
[arXiv:2208.08602 [hep-ph]].

\bibitem{Ellis:2004uz}
J.~R.~Ellis, M.~Karliner and M.~Praszalowicz,
JHEP \textbf{05}, 002 (2004)
doi:10.1088/1126-6708/2004/05/002
[arXiv:hep-ph/0401127 [hep-ph]].

 \bibitem{Yang:2015era}
  G.~S.~Yang and H.-Ch.~Kim,
  Phys.\ Rev.\ C {\bf 92}, 035206  (2015)
  [arXiv:1504.04453 [hep-ph]].
  


\bibitem {Cheng:2006dk}H.~Y.~Cheng and C.~K.~Chua,
Phys.\ Rev.\ D \textbf{75}, 014006 (2007) and
Phys.\ Rev.\ D \textbf{92}, 074014 (2015).

\bibitem{deSwart:1963pdg}
J.~J.~de Swart,
Rev. Mod. Phys. \textbf{35}, 916-939 (1963)
[erratum: Rev. Mod. Phys. \textbf{37}, 326-326 (1965)]
doi:10.1103/RevModPhys.35.916

\bibitem{Kaeding:1995re}
T.~A.~Kaeding and H.~T.~Williams,
Comput. Phys. Commun. \textbf{98}, 398-414 (1996)
doi:10.1016/0010-4655(96)00085-9
[arXiv:nucl-th/9511025 [nucl-th]].

\bibitem{Praszalowicz:2018upb}
M.~Praszalowicz,
Eur. Phys. J. C \textbf{78}, no.8, 690 (2018)
doi:10.1140/epjc/s10052-018-6173-6 

\end{thebibliography}
\end{document}